\documentclass[pra,twocolumn,showpacs,preprintnumbers,amsmath,amssymb,a4paper]{revtex4-1}
\usepackage{graphicx}
\usepackage{color}
\usepackage{SIunits}
\usepackage{units}
\begin{document}

\title{Excitability in semiconductor micro-ring lasers: experimental and theoretical pulse characterization}

\author{L. Gelens$^{1}$, L. Mashal$^{2}$, S. Beri$^{1,2}$, W. Coomans$^{1}$, G. Van der Sande$^{1}$, J. Danckaert$^{1,2}$, G. Verschaffelt$^{1}$}
\affiliation{
$^1$Department of Applied Physics and Photonics, Vrije Universiteit Brussel,
Pleinlaan 2, 1050 Brussels, Belgium;\\
$^2$Department of Physics, Vrije Universiteit Brussel,
Pleinlaan 2, 1050 Brussels, Belgium;
}

\date{\today}


\begin{abstract} 
We characterize the operation of semiconductor micro-ring lasers in an excitable regime.
Our experiments reveal a statistical distribution of the characteristics of noise-triggered optical pulses that is not observed in other excitable systems. In particular, an inverse correlation exists between the pulse amplitude and duration.
Numerical simulations and an interpretation in an asymptotic phase space confirm and explain these experimentally observed pulse characteristics.
\end{abstract}

\maketitle

\section{Introduction}

Excitability --- the ability of nonlinear systems to fire large well-defined
output pulses when a threshold is crossed --- has been investigated in
a broad range of scientific areas, including physics, chemistry, biology and neuroscience \cite{Hodgkin52a,Kapral95a,Bertram95a, Maass97a,
Pikovsky97a,Izhikevich00a,Lindner04a}.
The strong interdisciplinarity of excitable systems has attracted the attention of theoreticians interested in their universal properties.
These theoretical studies disclosed that excitability takes place when a {\it separatrix} is crossed in the phase space of the system, and led to a comprehensive classification of excitable systems based on bifurcations taking place in their phase space \cite{Izhikevich00a,Bertram95a}. 
Considerable theoretical insight was further gained into the role played by networks of excitable units in neuroscience \cite{Rabinovich06a}. 
Moreover, the investigation of excitable units is of high practical relevance as networks of spiking neurons have proved to be computationally superior to other neural networks \cite{Maass97a}.

In the field of photonics, excitability is widely studied both theoretically and experimentally \cite{Yacomotti94a,Dubbeldam97a,Wieczorek02a,Gomila05a,Marino05a,Wunsche02a,
Giacomelli_PRL_2000,Goulding07a,Romariz07a}
and a broad search was started for optical excitable units that could be deployed in optical neural networks. 
In the last decade, lasers with saturable absorber
\cite{Yacomotti94a,Dubbeldam97a}, optically injected lasers
\cite{Goulding07a,Wieczorek02a}, lasers with optical feedback
\cite{Giacomelli_PRL_2000} or VCSELs with opto-electronic feedback
\cite{Romariz07a} have all been proposed as optical excitable units.

Most often excitable behavior in optical systems has been shown to occur close to a fold bifurcation and a homoclinic bifurcation of a {\it stable} limit cycle \cite{Goulding07a,Wieczorek02a,Yacomotti94a,Dubbeldam97a,Gomila05a}. 
In these systems, the response to noise is governed by the presence of an {\it accessible} saddle point $S$ embedded in the separatrix. Pulses are activated by noise-induced fluctuations which connect the resting state to $S$ and the large deterministic excursion takes place along the {\it unstable} manifold of $S$. In this scenario, excitability is therefore possible in the limit of vanishing noise intensity. 

In a recent paper \cite{Beri_PLA_2009}, we proposed that excitability can take place in systems with weakly broken Z$_2$-symmetry, which includes optical units such as semiconductor micro-ring lasers \cite{SorelOL2002, HillNature2004} and micro-disk lasers \cite{Liu_NatPhot_2010}. It can be shown that in this class of systems the accessible saddle is not embedded in the separatrix for excitability. Therefore, the unstable manifold of $S$ does not participate in excitability and the generation of pulses cannot be initiated in the vanishing noise limit. 
Clear evidence of such excitability was shown in semiconductor ring lasers with weakly broken Z$_2$-symmetry in the presence of a non-vanishing, finite intensity of spontaneous emission noise \cite{Beri_PLA_2009}.

In this paper, we address that micro-ring lasers (MRLs) in the excitable regime show a notable degree of variation in the amplitude and width of the excited pulses. We characterize both theoretically and experimentally the particular pulse properties when the triggering occurs through optical noise in the system. The main result is that a clear inverse correlation is observed between the amplitude and width of the excited pulses. In Section \ref{Sect::experiments}, we experimentally investigate the excitable behavior of a MRL in the time domain. 
A characterization of the stochastic properties of excitability is carried out to reveal an inter-spike-interval (ISI) distribution which is exponential for large waiting times, but diverges from the Kramers form for intervals below approx.\ 50 ns.
In Sections \ref{Sect::models} and \ref{Sect::stochAS}, we use a general rate-equation model and an asymptotic model for excitability in MRLs as introduced in Ref.\ \cite{Beri_PLA_2009, Gelens_EPJD_SRL_2010} to explain the 
experimentally observed features. 
Finally, in Section\ \ref{Sect::conclusion}, we discuss the main results presented and their generality.

\begin{figure}[t!]
\centering
\includegraphics[width=8 cm]{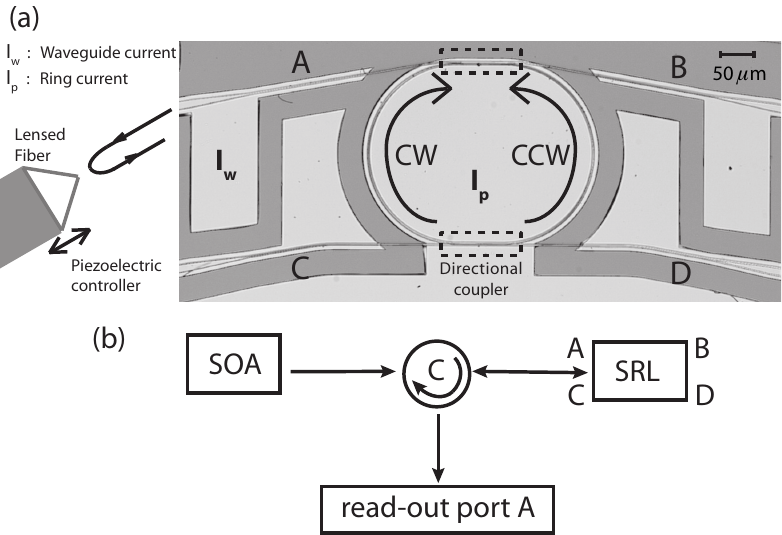}
\caption{\label{Fig:set_up} Experimental MRL set-up. (a) depicts the actual MRL device with its contacting.
The ring itself is pumped with a current $\text{I}_{\text{p}}$. Four waveguide contacts are depicted of which only port A is biased with $\text{I}_{\text{w}}$. (b) shows the overall setup where the read-out and the SOA are separated from each other by a circulator C.
}
\end{figure}

\section{Experiments}\label{Sect::experiments}

\subsection{Device and Setup}

The experiments have been performed on an InP-based multi-quantum-well MRL with a racetrack geometry [see Fig.~\ref{Fig:set_up}(a)]. 
The optical power is coupled out of the ring cavity by directional
coupling to bus waveguides which are integrated on the same optical chip. 
The use of two bus waveguides in the micro-ring design allows for four
independent input/output ports A-D which can be accessed with optical fibers.
The device is mounted on a copper mount and is thermally controlled by a Peltier
element which is stabilized with an accuracy of $0.01^{\circ}$C.

The MRL chip is deployed in a setup as shown in Fig.~\ref{Fig:set_up}.
In our experiments only port A of the four ports is used for
collecting output power from the MRL as well as injecting optical noise in the
micro-ring cavity.
We use a directly biased semiconductor optical amplifier [SOA in Fig.~\ref{Fig:set_up}(b)] operating in
the C-band to generate a controllable amount of noise through amplified spontaneous emission
which is injected in the ring through a lensed optical fiber coupled to port A
of the MRL.
The use of a circulator [C in  Fig.~\ref{Fig:set_up}(b)] allows us to read output from port A with the same
lensed fiber.

Electrical contacts have been applied to the bus waveguides that can be
independently pumped. They allow to amplify on-chip the signal emitted by
the ring. More interestingly, the presence of a contact allows us to
continuously break the symmetry of the device in a controlled way by making both the strength and the phase of the linear coupling between the CW and the CCW mode asymmetric.
Using the fiber's facet
as a mirror, we are able to reflect power from one mode (e.g.\ CCW) back into the
waveguide and finally to the counter-propagating mode in the ring. The application of a direct bias current $I_w$ on the 
waveguide's electrode has two main effects. First, it controls the power that is coupled to the CW mode; second, it affects the optical length of the waveguide via carrier-induced refractive index changes. Therefore, the phase of the reflected signal can be controlled by tuning the bias current on the waveguide and changing the position of the fiber facet which is piezo-controlled.
We have previously demonstrated in Refs.\  \cite{Beri_PLA_2009, Gelens_EPJD_SRL_2010} that such a breaking of the
Z$_2$-symmetry of the micro-ring leads to excitable behavior when the
optical noise intensity is large enough to make one of the states metastable.
We analyze the output power
of the CCW mode (port A) using a fast photodiode (with a bandwidth of 2.4 GHz) connected to an oscilloscope (with a maximum sampling rate of 20 Gs/s).

\subsection{Excitability}

\begin{figure}[t!]
\centering
\includegraphics[width=8 cm]{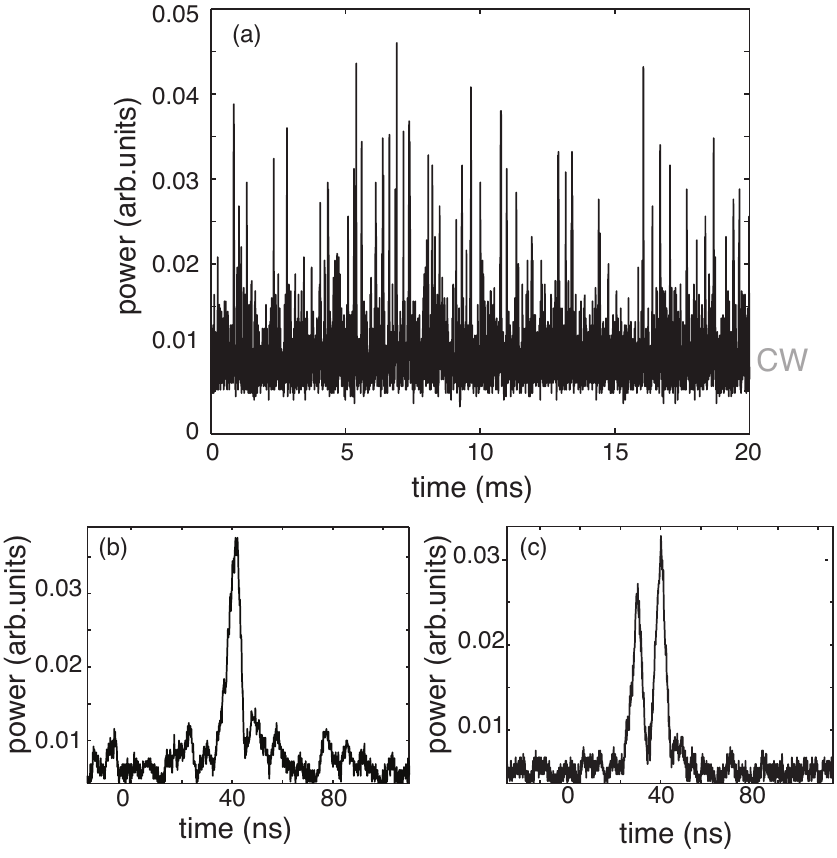}
\caption{\label{Fig:Exp_Excitability} 
(a) Experimental time series of port A demonstrating the excitable behavior of the MRL operating at $I_{p} = 47.45 $mA. The current on the waveguide of port A is $I_{w}=13$mA and the current in the external SOA is $I_{SOA} = 700 mA$. 
(b) and (c) show examples of a single and a double excited pulse taken from the same time series.}
\end{figure}

In the temperature range of operation, the transparency current density of our semiconductor material is $\sim 1.0\unit{kA/cm^{2}}$, which leads to a transparency current of $\sim 24 \unit{mA}$ for the MRL and  $\sim 4 \unit{mA}$ for the waveguide. The MRL device reaches threshold at $35 \unit{mA}$ and excitability is observed between $42 \unit{mA}$ and $48 \unit{mA}$.
A typical time series revealing excitable behavior of the micro-ring is
shown in Fig.~\ref{Fig:Exp_Excitability}(a). 
The device operates most of the time  in the CW unidirectional mode. However, pulses in the CCW direction can be regularly observed. 
An example of an excited pulse with a pulse duration of $\sim 7$ns is shown in Fig.~\ref{Fig:Exp_Excitability}(b).

In contrast to what is expected for a typical excitable system, not all
excited pulses have the same amplitude and duration, see e.g.\ Fig.~\ref{Fig:Exp_Excitability}(a). We clearly observe a degree of variation in the amplitude of the pulses (which is not a consequence of undersampling since it persists for higher sampling rates).
In the same way, a distribution of pulse-durations is observed in the
experimental time series and the pulse-durations seem to be spread between a
minimum of $7$ns and a maximum of $20$ns. Such a spread is significantly larger
than the sampling time.

\begin{figure}[t!]
\centering
\includegraphics[width=8 cm]{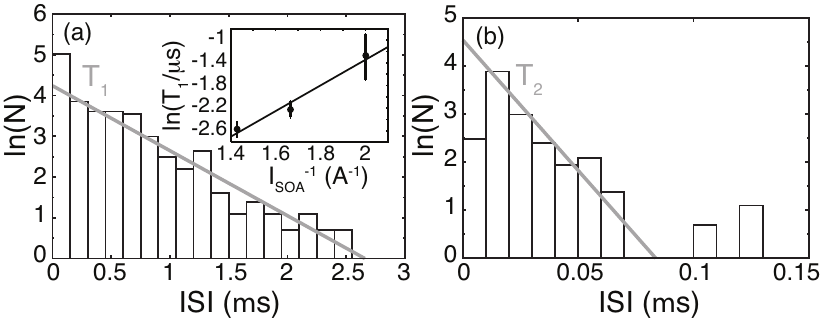}
\caption{\label{Fig:Exp_ISI}
Inter-Spike-Interval (ISI) distribution for pulses emitted by the symmetry
broken MRL for device parameters as given in Fig. \ref{Fig:Exp_Excitability}. The logarithm of the number of events $N$ in each ISI bin is shown. (a) The distribution of 'long' ISIs is well-fitted by an
exponential curve (see gray solid line). The gradient $1/T_1= 1.6 \mu$s$^{-1}$
indicates a characteristic time-scale of $\sim 0.63$~$\mu$s. 
(b) The distribution of 'short' ISIs reveals a different time constant $T_2 \sim 18.4$ns. The inset in panel (a) shows the dependence of the 'long' ISI time-scale $T_1$ (with error bars given by the standard deviation from the mean value $T_1$) on the current on the external SOA ($I_{SOA}$) which is used as noise source.}
\end{figure}

\subsection{Inter-spike-interval distribution}

In order to quantitatively describe the ISI, we have constructed the
ISI-distribution from the experimental time-series. A threshold is put at 60 \% of the maximum pulse height recorded in the time series. Only pulses above this threshold are taken into account in the calculation of the ISI distribution. We have verified that calculating the ISI distribution for other values of the threshold gives the same qualitative results as long as it is not chosen too low as one will then start to measure noisy excursions instead of actual excited pulses.
An example of such an ISI-distribution is
shown in Fig.~\ref{Fig:Exp_ISI}(a) for the MRL operating at a current of $47.45$
mA.
The tail of the distribution is well fitted by a typical Kramers' exponential distribution $P
\left( \tau \right) \propto \exp \left[ -\tau / T_1\right]$, which
indicates that the excited pulses are activated by noise-induced crossing of a
potential barrier.
The time constant $T_1$ obtained from the fitting of
Fig.~\ref{Fig:Exp_ISI}(a) is $T_1 = 0.63$ $\mu$s. 

A significant deviation from the Kramers' law
is, however, observed for ISIs shorter than $50$ ns, as shown in Fig.~\ref{Fig:Exp_ISI}(b).
The distribution of such 'short' ISI is strongly peaked around approx.\ 20ns and decreases
abruptly with a characteristic time $T_2 = 18.4$ns. 
A similar deviation from the Kramers' law was previously reported
in the residence-time distribution of MRLs \cite{Beri_PRL_2008} and in other
optical excitable systems such as lasers with optical feedback close to the low frequency fluctuations
regime \cite{Yacomotti99a, EguiaPRE2000}.

In order to confirm and quantify the noise-activated origin of excitability, we have
investigated the dependence of the average Inter-Spike-Interval 
$T_1$ as function of the bias current on the ring $I_p$ and the
external noise.
The external noise source is the external semiconductor optical amplifier [see Fig.\ \ref{Fig:set_up}(b)] and it is biased at different DC currents. We assume a linear
relationship between the SOA current and the noise intensity coupled
into the MRL.
In the inset of Fig.~\ref{Fig:Exp_ISI}(a) the dependence of $T_1$
versus the SOA current is shown. The linear relation between $\log(T_1)$ and the inverse of the SOA current is consistent with a Kramers rate across a potential barrier
\cite{Melnikov91a}.
We have also verified that $T_1$ increases with the bias current $I_p$ on the ring, which is consistent with
the theoretical result that the activation energy of the MRL increases with the bias current \cite{Beri_PRA_2009}.

In contrast,  we have checked that the `short' ISI has a much smaller dependence on the noise strength and remains approximately constant when changing the $I_{SOA}$ and $I_{p}$.
The origin of the non-Arrhenius distribution for these 'short' ISIs is instead related to deterministic dynamics. This will be discussed in more detail in Sections \ref{Sect::models}-\ref{Sect::stochAS}.

\subsection{Amplitude-width distribution} \label{subsec_exp_amp_width}

\begin{figure}[t!]
\centering
\includegraphics[width=8 cm]{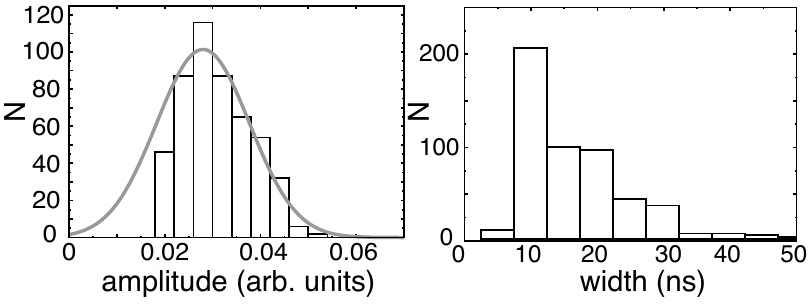}
\caption{\label{Fig:heigth_width_distr}
Distribution of (a) the pulse amplitude and (b)
the pulse width for $I_{p} = 47.45$mA, $I_{SOA} = 700$mA,
$I_{w}= 13.0$ mA. The number of events $N$ in each bin is plotted. The gray solid line shows the corresponding Gaussian distribution as predicted by the theory presented in Section \ref{Sect::stochAS}.}
\end{figure}

In order to further characterize the properties of the excited pulses, we
build the distribution of the pulse duration, defined as the 
full-width-half-maximum (FWHM) and the distribution of pulse amplitudes. 
Typical examples of the amplitude and width distributions are shown in
Fig.~\ref{Fig:heigth_width_distr}.

A distribution of pulse amplitudes is evident in
Fig.~\ref{Fig:heigth_width_distr}(a), which is consistent with the pulse amplitude
modulation present in Fig.~\ref{Fig:Exp_Excitability}. 
Such distribution of amplitudes was not observed in other optical excitable
systems  (see for instance
\cite{Kelleher09a,Yacomotti94a,Wunsche02a})
and is not consistent with a
standard excitability scenario where the system performs a large phase space
excursion following the unstable manifold of a saddle structure
\cite{Izhikevich00a,Lindner04a} for which a much sharper distribution
is expected. The width of the excited pulses is distributed asymmetrically around an
average value of $18$ ns and reveals a non-zero
probability for pulses as long as $40$ ns.

\begin{figure}[t!]
\centering
\includegraphics[width=8 cm]{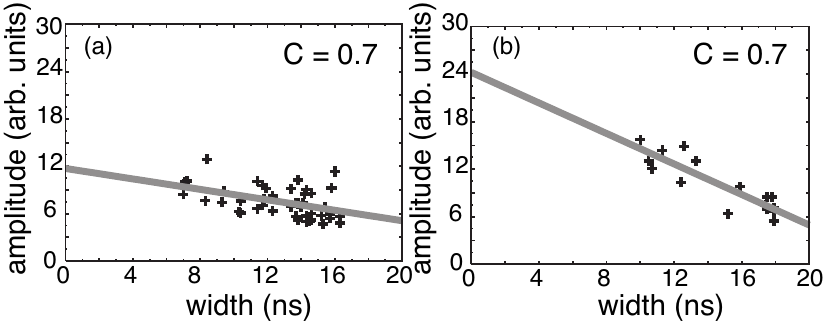}
\caption{\label{Fig:curves_h_w_exp} Measured width and
amplitude of the excited pulses for two different values of the bias current $I_p$ on
the ring. $I_{SOA} = 0 $mA and $I_{w}= 13.0$ mA. 
(a) $I_p = 42.82$mA and (b) $I_p = 44.49$mA. A linear fit of the data points is shown by the gray, solid line. The correlation coefficient $C$ is approx.\ 0.7 in both cases.}
\end{figure}

In order to shed light onto the dynamical origin of such distributions, we have
measured the correlation between pulse amplitude and
width for different values of the bias current on the ring. We minimize the noise-contributions to the pulse amplitude by performing this
measurement without the external SOA.
The results are shown in Fig.~\ref{Fig:curves_h_w_exp}
and quantified using a correlation coefficient $C$:
\begin{eqnarray}
C &=& \frac{\sum  \left(w_i-\bar{w}\right) \left( h_i-\bar{h}\right)}{N_{tot}
\sigma_h \sigma_w } \\
\sigma_h &=& \sqrt{\frac{\sum (h_i-\bar{h})^2 }{N_{tot}}} \ ; \ \sigma_w =
\sqrt{\frac{\sum (w_i-\bar{w})^2 }{N_{tot}}}
\end{eqnarray}

For all used bias currents there is a clear correlation between the width and
amplitude of the pulses. 
The experiments
reveal that pulses with a higher peak power are narrower, whereas pulses with a
lower peak power are wider. For $I_{SOA} = 0 $mA and $I_{w}= 13.0$ mA, we have measured a correlation $C$ of about 0.7. The linear fit between the amplitude and width is indicated by the gray, solid line.
We remark that this correlation shown in Fig.~\ref{Fig:curves_h_w_exp} has
been obtained by defining the pulse-width by its FWHM; we have verified that other ways to quantify
the pulse duration (not shown) lead to consistent results.

To understand the origin of this amplitude-width correlation of the excited pulses, we will numerically analyze the pulse properties in Section
\ref{Sect::stochAS}. To allow for a clear presentation of the results, in the next section, we shortly retake our approach towards modeling the time evolution of a single-longitudinal, 
single-transverse mode in a micro-ring cavity using a general rate-equation model for MRLs and an asymptotically reduced model. In particular, we explain the excitability scenario for MRLs in the asymptotic phase-space. For more information on the rate-equation model, we refer to Refs.\ \cite{VanderSandeJPhysB2008, Gelens_PRE_2009, Gelens_EPJD_SRL_2010}.

\section{Modeling and origin of excitability in semiconductor ring lasers}\label{Sect::models} 

In order to model the MRL operating in single-longitudinal and single-transverse mode, we use a rate-equation model for the 
the evolution of the slowly varying amplitudes of the counter-propagating modes $E_{cw, ccw}$ and the carrier inversion $N$
\cite{Gelens_EPJD_SRL_2010}:

\begin{eqnarray}
\frac{d{E}_{cw}}{dt} &=& \kappa(1+i\alpha)\left[g_{cw}N-1\right]E_{cw}  
\nonumber \\
&&- (k - \Delta k/2) e^{i (\phi_k - \Delta \phi_k/2)}E_{ccw} + \xi_{cw} \label{Eq::Field1::Original_ch7} ,\\
\frac{d{E}_{ccw}}{dt} &=& \kappa(1+i\alpha)\left[g_{ccw}N-1\right]E_{ccw}
\label{Eq::Field2::Original_ch7} \nonumber \\
&&- (k + \Delta k/2) e^{i (\phi_k + \Delta \phi_k/2)}E_{cw} + \xi_{ccw} ,\label{Eq::Field2::Original_ch7} \\
\frac{dN}{dt} &=& \gamma [ \mu -N - g_{cw}N|E_{cw}|^2  - g_{ccw}N|E_{ccw}|^2
]\label{Eq::Carriers::Original_ch7} 
\end{eqnarray}
 where $g_{cw} = 1-s|E_{cw}|^2-c|E_{ccw}|^2$, $g_{ccw} =
1-s|E_{ccw}|^2-c|E_{cw}|^2$ is a differential gain function which includes
 phenomenological self ($s$) and cross ($c$) saturation terms. 
$\kappa$ is
the field decay rate, $\gamma$ is the inversion decay rate, $\alpha$ is
the linewidth enhancement factor of the semiconductor material and $\mu$
is the renormalized injection current with $\mu\approx0$ at transparency and $\mu\approx1$ at lasing threshold. For the device such as the one discussed Sec.\ \ref{Sect::experiments}, $\mu=0$ corresponds to 25mA and $\mu=1$ to $35mA$, yielding the approximate relation $I_p \approx (10 \mu + 25)$mA.

\begin{table}
	\centering
		\begin{tabular}{lll}
		\textbf{Symbol} & \textbf{Physical meaning} & \textbf{Simulation value} \\
		\hline
		$\kappa$ & Field decay rate & $100 \unit{ns}^{-1}$\\
		$\gamma$ & Carrier inversion decay rate & $0.2 \unit{ns}^{-1}$\\
		$\alpha$ & Linewidth enhancement factor & $3.5$\\
		$\mu$ & Renormalized bias current & $1.65$\\
		$s$ & Self-saturation coefficient & $0.005$\\				
		$c$ & Cross-saturation coefficient & $0.01$\\				
		$k$ & Coupling amplitude & $0.44 \unit{ns}^{-1}$\\
		$\Delta k$ & Coupling amplitude asymmetry & $0.044 \unit{ns}^{-1}$\\		
		$\phi_k$ & Coupling phase & $1.5$\\
		$\Delta \phi_k$ & Coupling phase asymmetry & $0$\\
		\end{tabular}
	\caption{Summary of the physical meaning of the parameters in the rate equations (\ref{Eq::Field1::Original_ch7})-(\ref{Eq::Carriers::Original_ch7}) and their typical values used throughout this article, unless stated otherwise.}
	\label{tab:parametersSRL}
\end{table}

Asymmetric linear coupling terms are present in
Eqs.~(\ref{Eq::Field1::Original_ch7})-(\ref{Eq::Field2::Original_ch7}) between $E_{cw}$ and $E_{ccw}$, which model a
backscattering of power from one mode to the other.
Such intrinsic backscattering originates from reflections at the directional coupler or
at the chip facets, and is in general asymmetric
due to unavoidable imperfections introduced during
device fabrication. Moreover, asymmetries in backscattering introduced externally in our set-up (see Fig.\ \ref{Fig:set_up}), such as reflections at the fiber tip at one side of the chip, are also lumped into these coupling terms. The parameters $k$ and $\phi_k$ represent respectively the average coupling amplitude and phase, whereas the coupling asymmetry is described by the symmetry breaking terms $\Delta k$ and $\Delta \phi_k$. Unless mentioned otherwise, throughout this manuscript we will use the parameters shown in Table~\ref{tab:parametersSRL}.

Noise terms are introduced in
Eqs.~(\ref{Eq::Field1::Original_ch7})-(\ref{Eq::Carriers::Original_ch7})
as complex, Gaussian, zero-mean stochastic terms  $\xi_{cw,ccw}$ described by
the correlation terms $\langle
\xi_i(t+\tau) \xi^*_j(t) \rangle=2DN\delta_{ij}\delta(\tau)$, where $i,j=$\{cw,
ccw\} and $D$ is the noise intensity \cite{Perez_noise_2009}. 

All simulations are performed in this rate equation model (\ref{Eq::Field1::Original_ch7})-(\ref{Eq::Carriers::Original_ch7}). However, we will interpret our results in a two-dimensional phase-space corresponding to a reduced MRL model that has been proven to be a powerful tool to model the slow time dynamics in MRLs \cite{Beri_PRL_2008,Gelens_PRL_2009,Beri_PLA_2009, Gelens_EPJD_SRL_2010}. This two-dimensional model is valid on time scales slower than those of the relaxation oscillations. The two variables $\theta \in[-\pi/2,\pi/2]$ and $\psi\in[0,2\pi]$ are defined by
\begin{subequations}
\label{eq:RedEqDef}
\begin{align}
\theta &\equiv 2\arctan\left(\frac{|E_{cw}|}{|E_{ccw}|}\right) - \frac{\pi}{2},\\
\psi &\equiv \angle E_{ccw} - \angle E_{cw}.
\end{align}
\end{subequations}
$\theta$ is a measure for the power partitioning between the counter-propagating modes and $\psi$ is the phase difference between the corresponding electric fields. 

\begin{figure}[t!]
\centering
\includegraphics[width=8 cm]{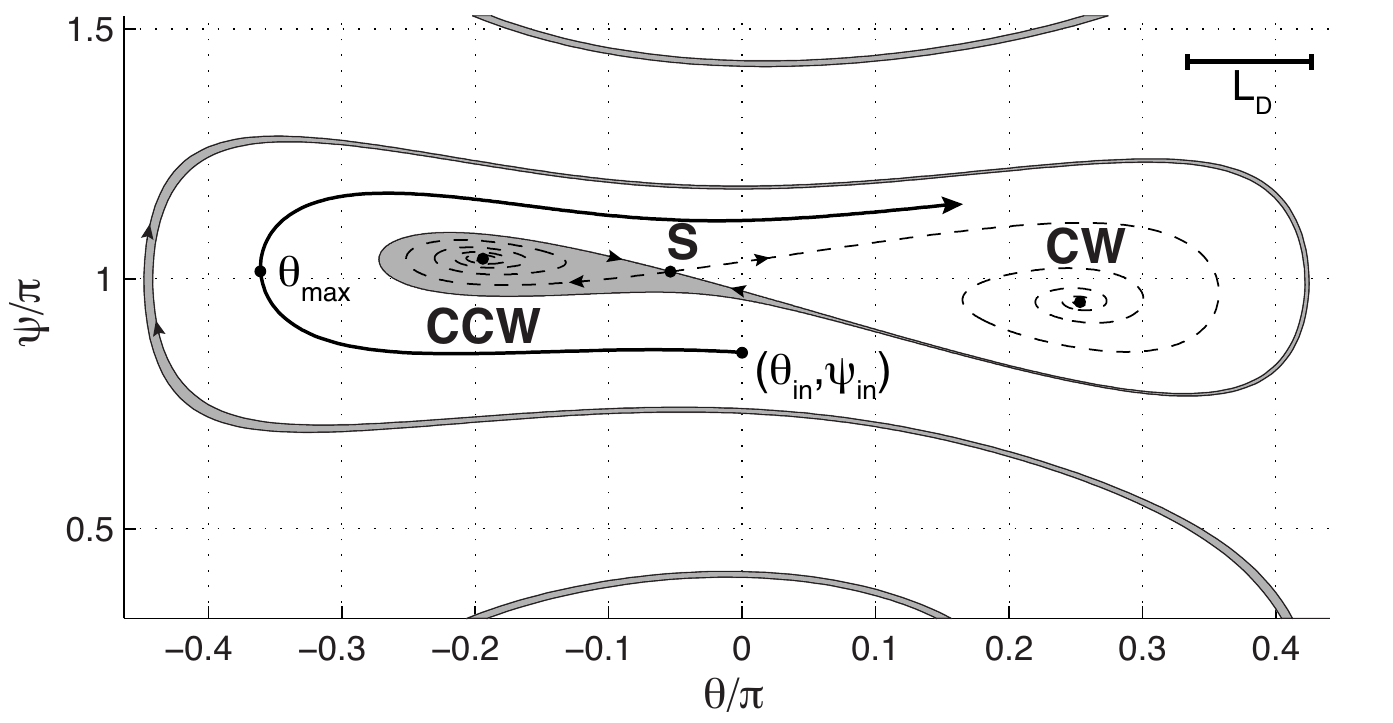}
\caption{\label{fig:ExcPhSp} 
The phase space topology of an asymmetric MRL. CW,
CCW and S are respectively two quasi-unidirectional states and a saddle state of
which the stable and unstable manifolds are displayed. The basin of attraction
of the CW (CCW) quasi-unidirectional state is colored white (gray). Parameters are chosen as in Table~\ref{tab:parametersSRL}.}
\end{figure}

To describe the excitability scenario in MRLs, we consider a ring laser operating unidirectionally whose symmetry has been slightly broken. 
The corresponding phase space is shown in Fig. \ref{fig:ExcPhSp}.
Two counter-propagating
unidirectional stable attractors are present in the MRL ---the CW and the CCW
mode--- and are depicted here in the two-dimensional $(\theta,\psi)$
phase space. The white and gray regions indicate the
basins of attraction of the CW and the CCW mode. They are separated by the
stable manifolds of a saddle state indicated by S. 

The presence of an asymmetry parameter manifests itself in
the different stability of the modes, leading to different sizes of their basins
of attraction. 
In particular, the distance between two branches of the stable manifold of $S$
is affected by the symmetry breaking and can be made arbitrarily
small by controlling the parameters of the system \cite{Gelens_EPJD_SRL_2010}.
When noise is present in the system, a diffusion length-scale $L_D$ appears,
which depends on the noise intensity $D$.
The onset of excitability in MRLs is regulated by the interplay between $L_D$
and the distance between the folds of the stable manifold of $S$.

Assume that the MRL is operating in a regime such that $L_D$ is small
compared to the size of the basin of attraction of CW
and large compared to the distance between the branches of the stable manifold
of $S$.  
The MRL will spend most of the time in the vicinity of the CW stable
state; however a rare large fluctuation may move the system to the boundary
of the basin of attraction of CW. When this happens, the system will cross this boundary
with an overwhelming probability by crossing  {\it both branches} of the stable manifold of $S$
and thereafter perform a large deterministic excursion leading to the emission
of a CCW pulse.

This excitability scenario is different from the one most frequently
encountered in optics, where pulses are initiated stochastically by
crossing only {\it one branch} of the
stable manifold of a saddle and completed largely deterministically by following a
branch of the {\it unstable} manifold of the same saddle back to the initial
quiescent
state \cite{Yacomotti94a,Dubbeldam97a,Wieczorek02a,Wunsche02a,Goulding07a}. Even in the rare situations where the system was shown to be both bistable and excitable at the same time, the excitable excursions are still completed by following the unstable manifold of the saddle point \cite{Gelens_PRA_2008}. 
In MRLs, however, this scenario is forbidden by the residual Z$_2$-symmetry and the
unstable manifold of $S$ which connects with the metastable CW state
\cite{Beri_PLA_2009}. 

This specific mechanism of excitability is general for all systems with weakly broken Z$_2$-symmetry and occurs near a homoclinic bifurcation that unfolds from a Takens-Bogdanov point. In that homoclinic bifurcation an unstable cycle is created, which later disappears in a fold of cycles \cite{Gelens_EPJD_SRL_2010}. Such a sequence of bifurcations leads to the folded shape of the stable manifold of the saddle (see Fig.\ \ref{fig:ExcPhSp}), necessary for the system to be excitable. The unfolding of the different bifurcations from a Takens-Bogdanov point has been characterized in depth, both in systems with Z$_2$-symmetry \cite{Knobloch_1981, Dangelmayer_Hopf_1991} and in systems where this symmetry is broken \cite{DangelmayrGuckenheimer}. As an example of other systems that share the same symmetry as the MRL and which can therefore exhibit similar excitable behavior when weakly breaking this symmetry, we mention e.g.\ CO$_2$-lasers \cite{Angelo_PRL_1992} and oscillatory convection in binary fluid mixtures \cite{Dangelmayer_1991}. Perhaps the most obvious example with the same circular symmetry as MRLs are semiconductor micro-disk lasers \cite{Liu_NatPhot_2010}. In semiconductor micro-disk lasers where the two lasing modes are the whispering gallery modes, the appropriate rate-equation model is identical to the one studied in this work.

\section{Stochastic analysis and comparison to the experiments}\label{Sect::stochAS}

In this Section, we use direct numerical integration of
Eqs.~(\ref{Eq::Field1::Original_ch7})-(\ref{Eq::Carriers::Original_ch7}) using a stochastic Euler-Heun method to characterize the specific features of excited pulses in MRLs. More specifically, numerical time-series are collected and statistics of the ISI, the pulse amplitude and the pulse width are built. A projection of the time-series on the 
reduced phase space $(\theta,\psi)$ is then performed in order to validate the topological arguments of Sec.~\ref{Sect::models} explaining the spread in amplitude of the pulses and the associated correlated spread in their width, as also observed experimentally (see Figures\ \ref{Fig:heigth_width_distr}-\ref{Fig:curves_h_w_exp}).

\subsection{Pulse amplitude and width}

\begin{figure}[t!]
\begin{center}
\includegraphics[width=8 cm]{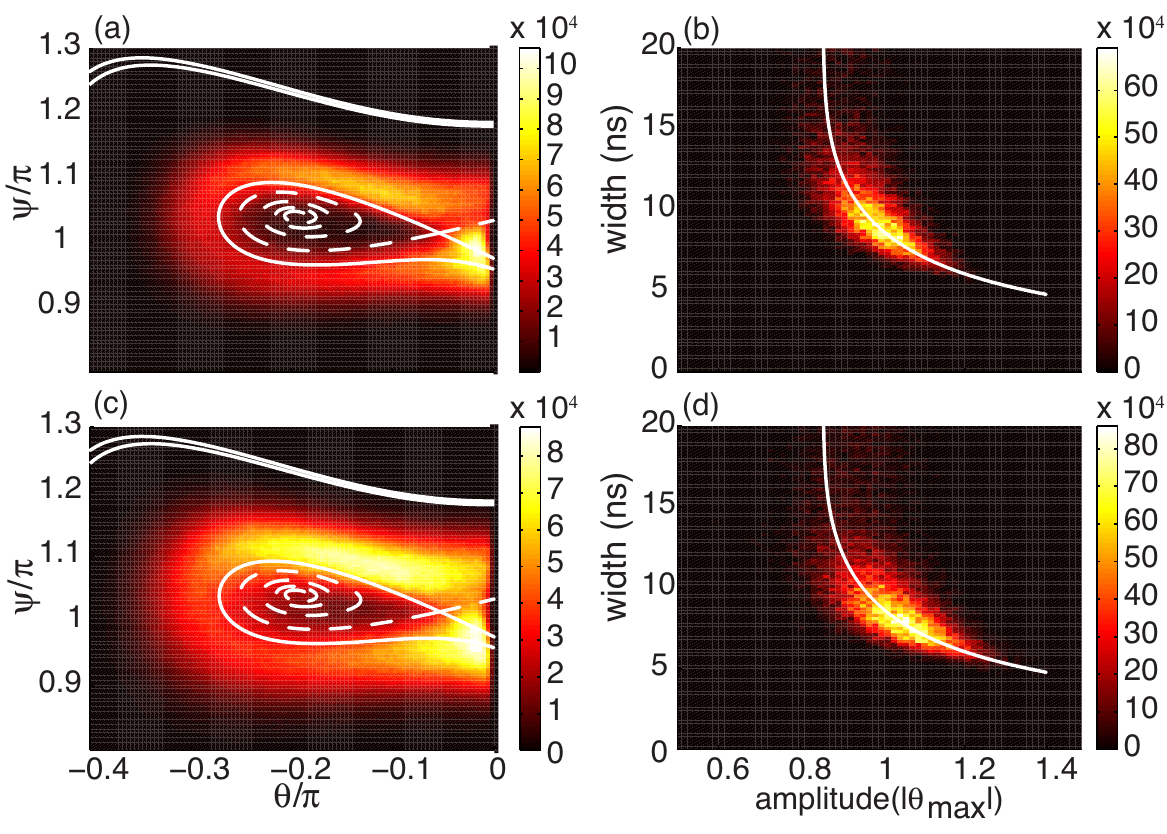}
\caption{(Color online) Results of simulations of Eqs.\ 
(\ref{Eq::Field1::Original_ch7})-(\ref{Eq::Carriers::Original_ch7}). (a)-(b):
$D=1.5\times10^{-4}$ ns$^{-1}$, $t_{obs}=100$ ms, (c)-(d): $D=2.5\times10^{-4}$
ns$^{-1}$, $t_{obs}=10$ ms, with $D$ the noise strength and $t_{obs}$ the observation time. (a) and (c): Histogram of the trajectories in asymptotically reduced two-dimensional
phase space. The full (dashed) white lines indicate the stable (unstable)
manifolds of the saddle. (b) and (d): Pulse width vs pulse amplitude. The white
curve indicates the prediction from the deterministic reduced model. Parameters are taken as in Table~\ref{tab:parametersSRL}.}
\label{fig:PulseWidthHeight}
\end{center}
\end{figure}

Fig. \ref{fig:PulseWidthHeight} shows the distribution of the 
phase space trajectories corresponding to excited pulses. 
In order to avoid sampling of double pulses, the system is reinitialized to the
original CW mode when the tail of the pulse satisfies the condition $\theta
>0$. Rare excursions that reside longer than 20 ns around the metastable CCW state are
discarded. Figs. \ref{fig:PulseWidthHeight}(a) and (c) show histograms of the
collected trajectories projected on the asymptotically reduced
two-dimensional
phase space ($\theta, \psi$), and this for two different values of the noise
strength $D$. The full (dashed) white lines indicate the stable (unstable)
manifolds of the saddle S. 

It is clear from Fig.~\ref{fig:PulseWidthHeight}(a) and (c) that
noise-activated trajectories cross the excitability threshold at a finite
distance from the saddle $S$.
Such saddle-avoidance is expected in stochastic systems where $L_D$ is not the
shortest length scale \cite{Luchinsky99a} and it is therefore compatible with
the mechanism described in Sec.~\ref{Sect::models}.

After crossing the stable manifolds of $S$, the trajectories spread in the
phase space due to diffusion. A distribution of pulse amplitudes and pulse widths as
experimentally observed in Figs.~\ref{Fig:Exp_Excitability} and
\ref{Fig:heigth_width_distr} can therefore be expected in MRLs as the
deterministic evolution of the pulse does not take place along a unique
trajectory. 
In the reduced phase space, pulses are a one-parameter family of trajectories that
can be parameterized by their initial conditions $\left( \theta_{in}
, \psi_{in} \right)$ [see Fig.~\ref{fig:ExcPhSp}].  By fixing $\theta_{in} = 0$ and $\psi_{in}$ beyond the
excitability separatrix, a sampling of excited pulses is achieved. The width of
each pulse can be quantified by the required time to return to the $\theta > 0$
condition; in the same way, the extreme value $\theta_{max}$ can be used to
quantify the pulse amplitude.

The correlation between amplitude and width can now be understood due to the
different velocity-fields at different positions in the reduced phase space. Higher (larger $| \theta |$) pulses move faster in
phase space and are thus narrower, while lower pulses are slower and consequently also wider.
The correlation curves extracted in such a way
are plotted in Fig.~\ref{fig:PulseWidthHeight}(b) and (d) (see the solid, white line). 
The observed  increase in pulse width with the decrease in pulse amplitude confirms the
experimental trend reported in Sec.~\ref{subsec_exp_amp_width}, i.e.\  we refer to Fig.\ \ref{Fig:curves_h_w_exp}.
A projection of the pulses, obtained by numerically solving the stochastic full-rate equation
model, in the amplitude-width space
[see point cloud in Fig.~\ref{fig:PulseWidthHeight}(b) and (d)] provides an extra confirmation of this deterministic prediction.

We observe that the theoretical amplitude-width curve is not evenly sampled by
the numerical pulses and that the numerical data clusters around a middle value. For this reason, we argue that the full profile of the amplitude-width curve
cannot be observed experimentally in our devices. The position of the point cloud in the amplitude-width space is slightly affected by noise.
The correlation relation holds for the two different noise
strengths, although it is clear that for lower noise strengths [Fig.
\ref{fig:PulseWidthHeight}(b)] the pulses tend to be less high in amplitude and
thus wider than in the case of higher noise strengths  [Fig.
\ref{fig:PulseWidthHeight}(d)]. 
The fact that the point cloud in Fig.~\ref{fig:PulseWidthHeight}(b) and (d) opens up for
decreasing pulse amplitudes is related to the nature of these pulses. These
relatively long pulses tend to wander around the metastable CW state or the
saddle point. Near the saddle point the deterministic trajectory greatly slows
down. 
Hence, in this case, the transit  times of the pulses are mainly determined by
the noise, giving rise to a larger spread.

\begin{figure}[t!]
\begin{center}
\includegraphics[width= 8 cm]{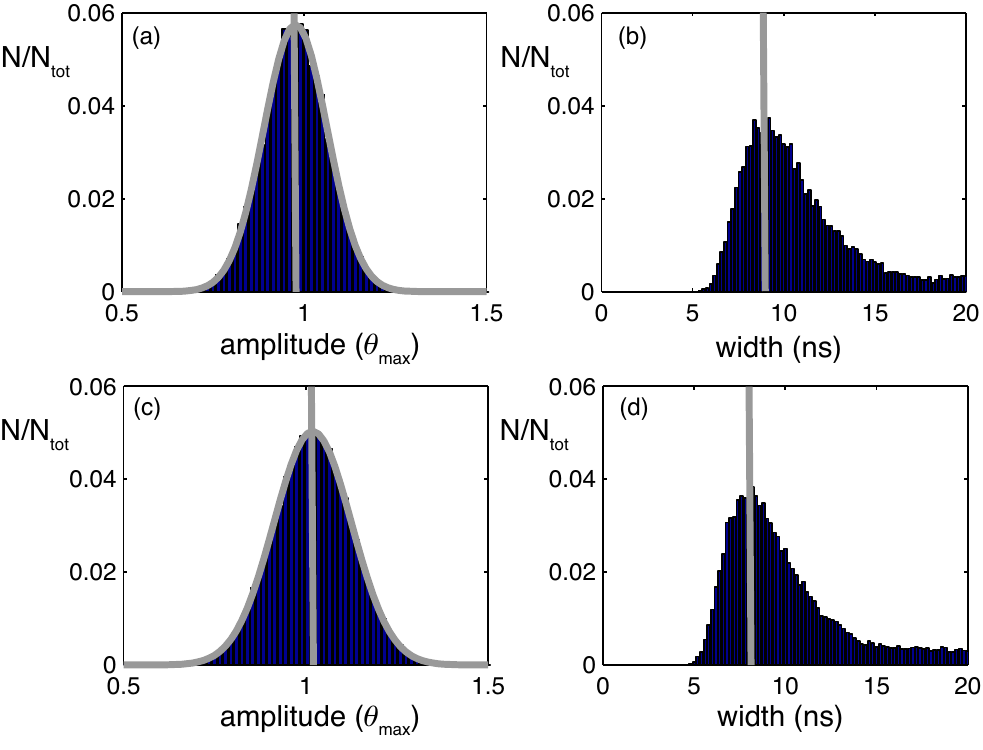}
\caption{(Color online) Results of simulations of  Eqs.\
(\ref{Eq::Field1::Original_ch7})-(\ref{Eq::Carriers::Original_ch7}). The histogram
of events $N/ N_{tot}$ is shown in function of the pulse amplitudes (a),(c) and in
function of the pulse widths (b),(d). $D=1.5\times10^{-4}$ for (a)-(b) and
$D=2.5\times10^{-4}$ for (c)-(d). The histogram of the pulse amplitudes can be
fitted by a Gaussian: $A\exp(-((x-B)/C)^2)$.  (a) $N_{tot} = $ 15052 events, $A=0.0575, B=0.9762, C=0.1237$.  (c) $N_{tot} = $ 23748 events, $A=0.0503,
B=1.019, C=0.1508$. The
vertical gray lines show the maxima of the histograms:  a width of $\approx 9.20$ ns is found in (b) and $\approx7.99$ ns in (d). Parameters are taken as in Table.\ \ref{tab:parametersSRL}.}
\label{fig:PulseHistGauss}
\end{center}
\end{figure}

A more tangible picture is given by considering the
width distribution and the amplitude distribution separately, instead of their mutual projection on the two-dimensional $(\theta,\psi)$ phase-plane. These histograms are shown in
Fig.\ \ref{fig:PulseHistGauss}. The pulse amplitude distribution can be properly
fitted by a Gaussian distribution $A\exp[-(\theta_{max}-B)^2/C^2]$, indicating
that the amplitude of the pulse is mainly determined by the magnitude of the
perturbation and less by the topology of the flow. The average pulse amplitude also
increases with increasing noise intensity, initiating excursions farther away
from the stable saddle manifold. In contrast, the pulse width distribution
is asymmetric. Some of the pulses tend to erratically wander around the
CCW state or the saddle giving rise to the tail in the pulse width distribution.
The average pulse width decreases with increasing noise intensity indicating
that the flow near the stable manifold is slower. This also confirms the pulse
amplitude-width trade-off trend of Fig. \ref{fig:PulseWidthHeight}. Similar experimental histograms have been shown in
Fig.\ \ref{Fig:heigth_width_distr}. Fig.\ \ref{Fig:heigth_width_distr}(a) showed that the spread in pulse amplitudes is consistent with a Gaussian distribution, while the spread in the width of the pulses in [see Fig.\ \ref{Fig:heigth_width_distr}(b)] is distributed asymmetrically.

\begin{figure}[t!]
\begin{center}
\includegraphics[width=8 cm]{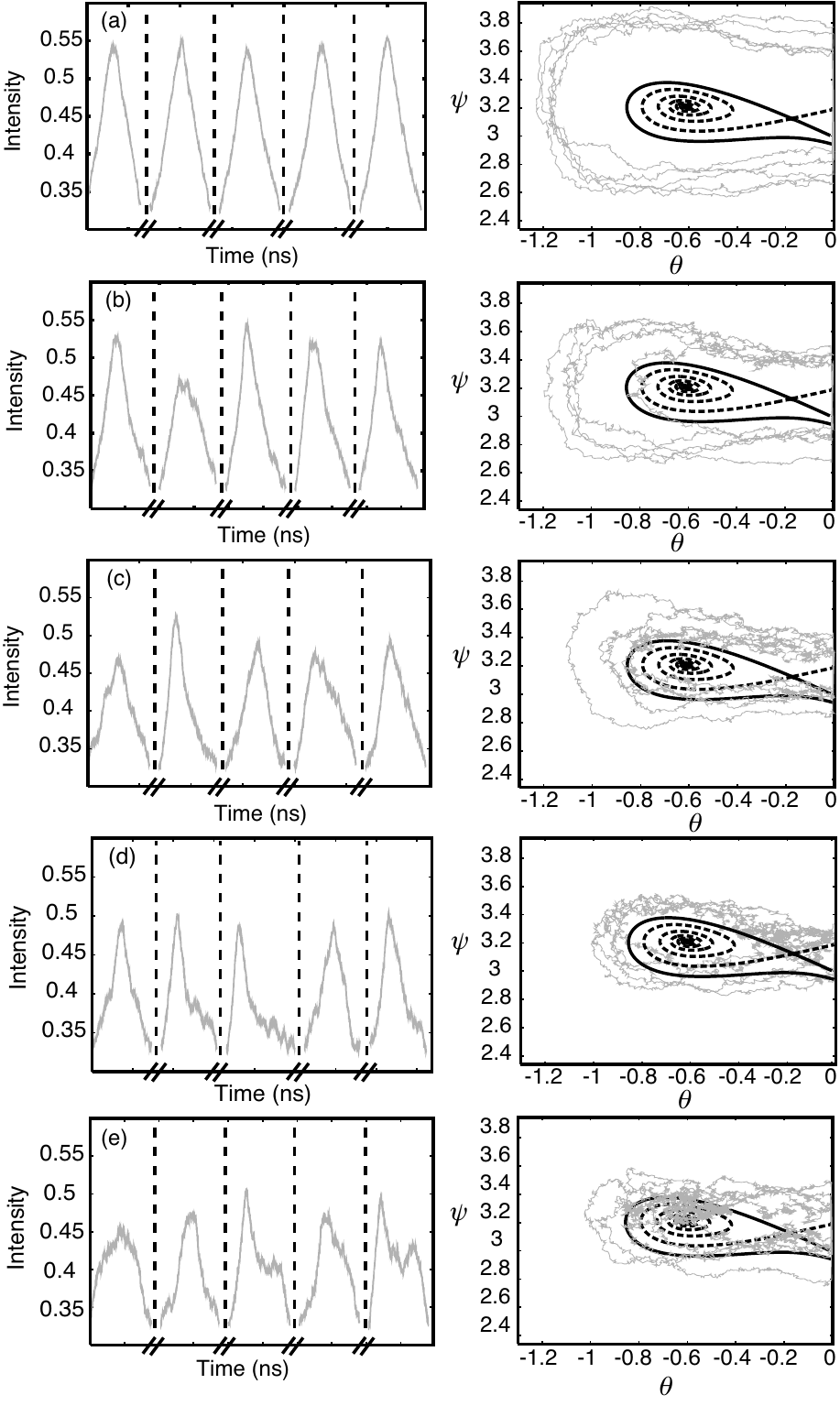}
\caption{Simulation of Eqs.\
(\ref{Eq::Field1::Original_ch7})-(\ref{Eq::Carriers::Original_ch7}). Typical
pulse shapes corresponding to different pulse widths are shown. Single pulses
are collected from the time traces in panels (a)-(e) with next to each panel
(a)-(e) the corresponding trajectory in the $(\theta,\psi)$ phase space in gray.
Stable (full) and unstable (dashed) manifolds are drawn in black. From (a) to
(e) pulses are shown with widths in the following intervals: 5-6 ns, 8-9 ns,
11-12 ns, 14-15 ns, 17-18 ns. $D=2.5\times10^{-4}$ ns$^{-1}$ and the other parameters are taken as in Table.\ \ref{tab:parametersSRL}.}
\label{fig:PulseShapes}
\end{center}
\end{figure}

Finally, in Fig. \ref{fig:PulseShapes} we show typical  pulse shapes as the
pulse width varies. Figs.\ \ref{fig:PulseShapes}(a)-(e) show a random collection
of pulses with widths in the following intervals: 5-6 ns, 8-9 ns, 11-12 ns,
14-15 ns, 17-18 ns, respectively. The corresponding trajectories in the
$(\theta,\psi)$ phase space are shown in gray. The faster, narrower pulses start
out at a relatively large distance from the stable manifold, and remain distant from it during the whole pulse trajectory. This is clearly visible in the pulses in Fig.
\ref{fig:PulseShapes}(a)-(b). Oppositely, pulses that start out closer to the
stable manifold slow down and are less high. For this type of excursions,
if the pulse trajectory comes too close to the stable manifold it can get
tangled up in the metastable CCW state or slowed down near the saddle, explaining
the formation of possible plateaus in the pulse. In the case one gets trapped in
the CCW state, the pulse would show a plateau at the pulse maximum, while if the
pulse slows down near the saddle, it would exhibit a plateau at the trailing edge of the
pulse.  In Fig. \ref{fig:PulseShapes}(d), many of the pulses come
very close to the saddle point, resulting in an even more pronounced slow-down, especially at the trailing edge of the pulse. Several rare pulse events where the system briefly gets stuck in the CCW state are depicted in panel (e). Also pulses that do not get close to the CCW state or the saddle point are characterized by an asymmetric pulse shape. Such asymmetry due to a slowing down of the pulse at the trailing edge (higher values of $\psi$) is best visible in panels (b) and (c).

\subsection{Inter-spike-interval diagram}

\begin{figure}[t!]
\begin{center}
\includegraphics[width= 8 cm]{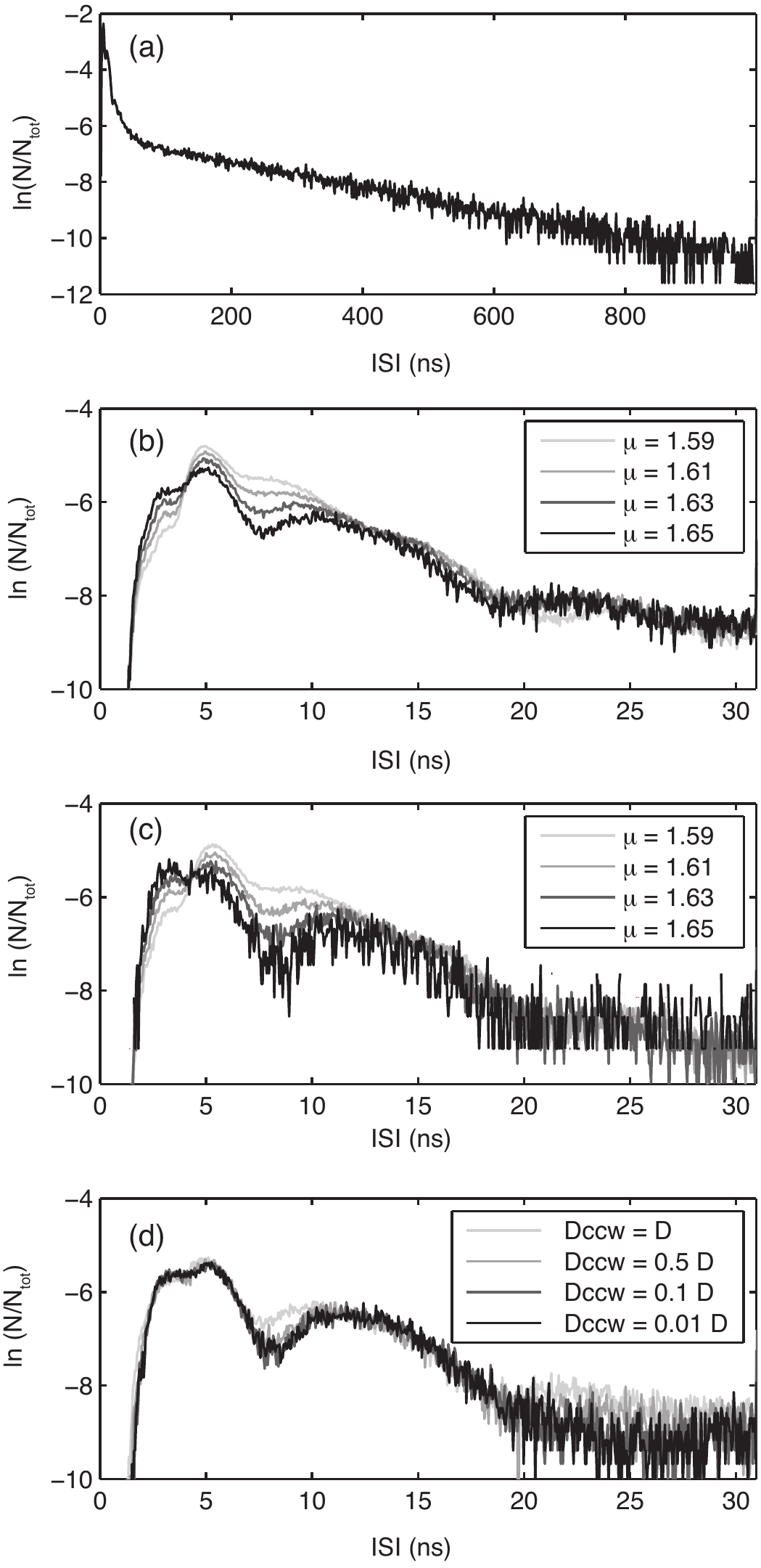}
\caption{Simulation of Eqs.\
(\ref{Eq::Field1::Original_ch7})-(\ref{Eq::Carriers::Original_ch7}) for $10 ms$ showing the
inter-spike-interval diagram. log$_2$(N/N$_{tot}$) is plotted in function of the
inter-spike-interval. In (a) $\mu = 1.65$, $D=2.5\times10^{-4}$ ns$^{-1}$ and the bin-size is chosen to be $1 ns$. In panels (b)-(d), a zoom of the peak at shorter ISI is shown with a bin-size of $0.05 ns$. Panels (b) and (c) show the variation of the ISI with $\mu$ for $D=2.5\times10^{-4}$ ns$^{-1}$ and $D=1.5\times10^{-4}$ ns$^{-1}$, respectively. Panel (d) demonstrates the change in ISI for asymmetric noise contributions to the $CW$ ($D_{cw} = D$) and the $CCW$ ($D_{ccw}$) mode for $\mu = 1.65$. The other parameters are taken as in Table.\ \ref{tab:parametersSRL}.}
\label{fig:histogram_ISI_log_N_Ntot_timemus}
\end{center}
\end{figure}

We have also determined the ISI distribution by simulating Eqs.\
(\ref{Eq::Field1::Original_ch7})-(\ref{Eq::Carriers::Original_ch7}) during a
time $t_{obs} = 10 \unit{ms}$. 
The distribution of the ISI between all excited pulses is shown in Fig.
\ref{fig:histogram_ISI_log_N_Ntot_timemus}. 
We have plotted the logarithm of the normalized number of events
log(N/N$_{tot}$) in function of the inter-spike-interval,
with $N$ the total amount of events in each bin and $N_{tot}$ the
total amount of excited pulses.

In Fig.\ \ref{fig:histogram_ISI_log_N_Ntot_timemus}(a), we have taken $\mu = 1.65$ and $D=2.5\times10^{-4}$ ns$^{-1}$ corresponding to the parameter set used before in Figures \ref{fig:PulseWidthHeight}-\ref{fig:PulseShapes}. The bin-size has been chosen to be $1$ ns. It becomes evident from Fig.
\ref{fig:histogram_ISI_log_N_Ntot_timemus}(a) that the total ISI distribution is the combination of two different time scales. The short ISIs and long ISIs are distributed in a different way and together represent a strongly Kramers'
type of behavior. 

The slow time scale can be fitted by an exponential curve [$\propto \exp{(-t / T_{1})}$], where the average ISI ($T_{1} $) is the fitting parameter. This slow time scale $T_{1}$ is typically in the order of $\unit{\micro s}$ and corresponds to the generation of the pulses in an Kramers' type noise-activation process across the excitability threshold \cite{Melnikov91a}. The short ISI times, however, are a signature of multiple consecutive excited pulses due to noise clustering. The possible excitation of double pulses such as those experimentally shown in Fig.~\ref{Fig:Exp_Excitability}(c) depend on the closeness of the stable and unstable manifolds of the saddle $S$. When these are close enough, noise can
excite a second pulse before the system can relax to the quiescent
state. The presence of such excited multi-pulses will show up in the ISI distribution as a sharp peak around the average pulse width. The numerical results in Fig.\
\ref{fig:histogram_ISI_log_N_Ntot_timemus}(a) confirm the experimental observation of both time scales in the experimental ISI distribution presented in Section \ref{Sect::experiments}  (see Fig.\ \ref{Fig:Exp_ISI}). 

The presence of
these time scales was also found 
in our study of stochastic mode-hopping in the bistable regime in Refs.\
\cite{Beri_PRL_2008, Gelens_PRL_2009, Beri_PRA_2009}. Both a mode-hop in the
bistable regime and an excitation beyond the excitability threshold are described
by a noise-activated escape, corresponding to the slow Arrhenius time-scale. The
fast non-Arrhenius character of the ISI finds its origin in a noise-induced diffusion
through both branches of the stable manifold, thus initiating another excursion
in phase space before relaxing to the CW state. Such a noise clustering of pulses due to the proximity of the relaxation trajectory of the excited pulse and the excitability threshold has been observed in several other noise driven excitable systems, such as e.g.\ lasers with optical feedback \cite{Yacomotti99a,EguiaPRE2000}, quantum-dot lasers with optical injection \cite{Kelleher09a,KrassiQD2010} and neurons of the Hodgkin-Huxley type \cite{Rowat2007}.

In panels (b)-(d), we study in more detail the ISI distribution of the faster time scale for different values of the pump current $\mu$, the noise strength $D$ and for asymmetric contributions of the noise to both counter-propagating modes. The bin-size is taken to be $0.05 \unit{ns}$.\\
Figs.\ \ref{fig:histogram_ISI_log_N_Ntot_timemus}(b)-(c) show the ISI distribution for varying values of the current $\mu$ at a fixed noise strength $D=2.5\times10^{-4}$ ns$^{-1}$ and $D=1.5\times10^{-4}$ ns$^{-1}$, respectively. One can notice that in all cases the maximum amount of events are located around $5 \unit{ns}$, which corresponds roughly to half of an excitable excursion\footnote{Defining the intensities of the CW and the CCW state by $P_{cw,ccw}$, we register an excited pulse from the moment the intensity in the CCW mode $P > P_{cw}+0.8(P_{ccw}-P_{cw})$. Likewise, the end of the pulse is defined as the moment when the intensity $P < P_{cw}+0.2(P_{ccw}-P_{cw})$. The ISI is calculated as the time between the onset of a pulse and the end of the previous pulse.}. For slightly higher ISI ($5 - 10 \unit{ns}$) a dip in the ISI distribution is observed. Such a dip is typical for noise clustering \cite{Yacomotti99a,Rowat2007} and is due to the nature of the relaxation trajectory of the excited pulse. In particular, it becomes evident from Fig.\ \ref{fig:histogram_ISI_log_N_Ntot_timemus}(b)-(c) that the dip becomes less pronounced for decreasing values of the current $\mu$ due to the fact that the relaxation to the stable node occurs increasingly slowly for lower values of $\mu$. One can also notice that the entire ISI curve moves up for decreasing $\mu$, which is a logical consequence of the decreasing depth of the potential well. It is interesting to note that the basin of attraction of the metastable CCW state decreases when decreasing $\mu$. In fact, similar excited pulses and ISI distribution still hold for $\mu = 1.59$ when the saddle and the metastable CCW state have disappeared. This can be understood by the fact that such dynamics arises as a scar of the bifurcations nearby \cite{Yacomotti99a}. 

All of the previous theoretical analysis has been done for the stochastic rate-equation system (\ref{Eq::Field1::Original_ch7})-(\ref{Eq::Carriers::Original_ch7}), where for reasons of simplicity the noise terms have been added in a symmetric way to both counter-propagating modes. From the experimental set-up explained in Section \ref{Sect::experiments}, one can wonder whether this symmetric assumption is valid as noise coming from the external SOA is injected mainly in one direction (the stable CW state). In order the check the validity of such an assumption, we have checked our analysis for asymmetric noise contributions, introducing noise with strength $D_{cw} = D$ in Eq.\ (\ref{Eq::Field1::Original_ch7}) for the stable CW mode and noise with strength $D_{ccw} < D$ in Eq.\ (\ref{Eq::Field2::Original_ch7}) for the metastable CCW mode. Qualitatively similar results are obtained as presented throughout this manuscript, which is reflected in Fig.\ \ref{fig:histogram_ISI_log_N_Ntot_timemus}(d) showing the ISI distribution for $\mu = 1.65$, $D_{cw} = D=2.5\times10^{-4}$ ns$^{-1}$ and different values of $D_{ccw}$.

\section{Discussion and Conclusion}\label{Sect::conclusion}

In conclusion, we have investigated the features of excitability in optical systems which are close to Z$_2$-symmetry and have elucidated the differences with other --- more common --- excitability scenarios \cite{Yacomotti94a,Dubbeldam97a,Wieczorek02a,Gomila05a,Marino05a,Wunsche02a,
Giacomelli_PRL_2000,Goulding07a,Yacomotti99a,Kelleher09a}.
The key message of this paper is the experimental and theoretical observation of a spread of the pulse amplitude and width in MRLs in the presence of noise, where a correlation between the amplitude and width of the excited pulses has been observed. Such a correlation between these two quantities is the signature of the deterministic evolution of the system once the separatrix is crossed.
Multi-pulse excitability \cite{Yacomotti99a} is also present due to the finiteness of the noise intensity and the relaxation to the quiescent state being slow due to the closeness to a homoclinic bifurcation in parameter space. 

For a realization of an all-optical neural network based on MRLs, such a pulse width distribution can present a drawback compared to other excitable systems. 
However, such drawback is eventually balanced by the easy integrability of MRLs on an optical chip and the possibility to control the coupling between ring cavity and waveguide \cite{Gabor_PTL_2009}.
Furthermore, excitability in MRLs does not require feedback from an external cavity or optical injection from a master laser as in the case of other optical systems.

\section*{Acknowledgements}
This work has been partially funded by the European Union under project
IST-2005-34743 (IOLOS). This work was supported by the Belgian Science Policy
Office under grant No.\ IAP-VI10. We acknowledge the Research
Foundation-Flanders (FWO) for individual support and project funding. Furthermore, we thank Gabor Mezosi and Marc Sorel for the fabrication of the semiconductor ring lasers on which the experiments have been performed and we thank Ingo Fischer for interesting discussions.


%

\end{document}